\documentclass{elsart5p}

\usepackage{natbib}
\usepackage{graphicx}
\usepackage{amssymb}

\begin{document}

\begin{frontmatter}

\title{The Interplay of Landau Level Broadening and Temperature
on Two-Dimensional Electron Systems}

\author{R. P. Gammag and C. Villagonzalo}
\ead{rayda.gammag@up.edu.ph}
\ead{cvillagonzalo@nip.upd.edu.ph}

\address{Structure and Dynamics Group, National Institute of Physics,\\
University of the Philippines, Diliman, Quezon City 1101, Philippines}

\begin{abstract}
This work investigates the influence of low temperature  and 
broadened Landau levels on the thermodynamic properties of two-dimensional electron 
systems. The interplay between these two physical parameters on the magnetic field 
dependence of the chemical potential, the specific heat and the magnetization is
calculated. 
In the absence of a complete theory that explains the Landau level broadening,
experimental and theoretical studies in literature perform different
model calculations of this parameter.
Here it is presented that different broadening parameters of Gaussian-shaped 
Landau levels cause width variations in their contributions to interlevel and intralevel excitations. Below a characteristic temperature, the interlevel excitations become negligible.
Likewise, at this temperature range, the effect of the Landau level broadening
vanishes.
\end{abstract}


\end{frontmatter}

\section{Introduction}
\label{intro}

Two-dimensional electron systems (2DES) are widely investigated due to their intriguing
non-bulk like properties and their possible applications \cite{AndoFS82}.
One example of interest is the oscillating behavior of their thermodynamic properties 
with respect to a strong perpendicular magnetic field $B$ at low temperature $T$
\cite{WildeSHH06,ZhuUMP03,WangTSS92,GornikLSS85}. This general trend is attributed to
the presence of Landau levels -- the discretized energy spectrum of a 2DES subject 
to a strong magnetic field applied normal to the 2D plane \cite{AndoFS82}. 
Landau levels are represented in the density of energy states (DOS) of a non-interacting 
2DES as a series of Dirac-delta functions with gaps that are empty of electronic states 
\cite{LiXD89}. 

On the side of theoretical investigations, 
earlier studies have obtained  Landau level broadening, in the extreme quantum limit, 
when self-consistent treatment of electron-impurity scattering is taken into account
\cite{DasSarma81a,DasSarma81b}.  
Another work, employing perturbative treatment of disorder and Coulomb interaction, 
obtained a DOS with sharp edges in the limit where Landau level mixing is ignored
\cite{MacDonaldOL86}.
Nevertheless, reports on the DOS extracted from experiments indicate broadened Landau 
levels. This is true whether the DOS is inferred from magnetization 
\cite{WildeSHH06,ZhuUMP03,EisensteinSNC85}, 
heat capacity \cite{WangTSS92,GornikLSS85}
or capacitance measurements \cite{MosserWKP86,SmithGSH85}.
Moreover, depending on the applied magnetic field strength, overlaps of Landau levels 
may occur which indicate existence of electronic states in between the ideal DOS peaks.
These overlaps have been analytically confirmed in the presence of a weak disorder 
\cite{DasSarmaX88}. But this derivation has no localization effects and it contains parameters related to the impurity configuration, namely, its density and distance from the 2DES plane. 
Such conditions make quantitative comparison with experimental results difficult.

Others resorted to phenomenological calculations employing either of these DOS shapes:
Gaussian \cite{EisensteinSNC85}, Lorentzian \cite{ZhuUMP03,OzdemirYO04},  
exponential \cite{OzdemirYO04} or Gaussian with a background 
\cite{WildeSHH06,ZhuUMP03,GornikLSS85}.
These depend on a single broadening parameter $\Gamma$, which 
according to the short range model is inversely proportional to $B^{1/2}$ \cite{AndoFS82}.
Several researchers have made data fits to obtain the DOS using this $B$ dependence 
on $\Gamma$ \cite{WildeSHH06,ZhuUMP03,MosserWKP86}.
However, the DOS obtained using $\Gamma\propto B^{1/2}$  is found to be four times 
smaller than measurements in 2DES in GaAs/AlGaAs heterostructures \cite{EisensteinSNC85}.
Furthermore, the Landau level widths are predicted to oscillate due to the
fluctuations in the screening of the disorder by the electrons in the 2DES
\cite{XieLD90,EsfarjaniGS90}. Experimental data from multiple-quantum-well 
structures are consistent with a $\Gamma$ oscillating as a function of $1/B$ \cite{WangTSS92}.
Although there is yet no consensus on the $B$-dependence of $\Gamma$,
it is still advantageous to use a broadening parameter as it contains 
all the collective temperature- and magnetic field-dependence of the 2DES.
This then makes comparison with experimental data of macroscopic properties easier.

Despite the lack of a complete theory for the Landau level broadening caused
by impurity disorder,
information on the effects of the broadening on the thermodynamic quantities
will provide further insights on the scattering mechanisms present.
In this work, the behavior of the specific heat and the magnetization as functions 
of $\Gamma$ and $T$ is obtained through a phenomenological approach. 
Here a Gaussian-shaped DOS is employed. 
This form is utilized since it is a commonly used model for experimental data and has
been found to fit better than other semi-elliptical functions and long-ranged models
\cite{WildeSHH06}. 
Using various broadening parameters, the above mentioned physical quantities are 
calculated for a single sub-band in temperatures in the order of 0.3 K to 4.2 K 
and magnetic fields in the order of 0.5 T to 15 T.

\section{The Simulation Model}

The specific heat capacity $C_V$ and the magnetization $M$ of  two-dimensional electron
systems are both functions of the magnetic field $B$ and temperature $T$. 
The specific heat at constant volume of an electron gas  is given as
\begin{equation}
C_V(B,T) = \frac{\partial}{\partial T} \int_{0}^{+\infty} f(E,\mu,T) (E - \mu) D(B,E) dE\;,
\label{eq:Cv}
\end{equation} 
where 
$f(E,\mu,T) = 1/(1+\exp\left[{(E-\mu)/k_BT}\right])$ is the Fermi-Dirac distribution function, 
$\mu=\mu(B,T)$ is the chemical potential, $k_B$ is the Boltzmann's constant,  
and $D(B,E)$ is the density of states. 

Similarly, the magnetization for an electron gas having no spin splitting
can be obtained from the free energy $F$,
\begin{equation}
M(B,T)=\left.-\frac{\partial F}{\partial B}\right|_{N=\mbox{constant}} \;,
\end{equation}
where 
\begin{equation}
F = \mu N-k_B T \int_{0}^{+\infty} D(B,E) \ln \left( 1 + \exp\left[\frac{\mu-E}{k_BT}\right] \right) dE,
\end{equation}
where $N$ is the electron concentration.
The behavior of these magneto-thermodynamic properties can be determined 
once the chemical potential and the density of states are known.

The  chemical potential $\mu$ can be obtained via a root-finding method for a fixed 
value of $N$. This condition is not a problem since 
experiments on 2DES usually keep $N$ constant.
For a given $T$ and  $B$, the value of $\mu$  is sought that provides the least
percentage error of not more than $0.001\%$ from the set value 
of the electron concentration in
\begin{equation}
N = \int_{0}^{+\infty} f(E,\mu,T) D(B,E) dE \;.
\label{eq:Concentration}
\end{equation}
In this work, $N$ is set equal to $3.6 \times 10 ^{11} \mbox{cm}^{-2}$. This value is
chosen to be of the same order of magnitude of high mobility 2DES samples as found 
in experimental reports \cite{ZhuUMP03,WangTSS92,GornikLSS85,EisensteinSNC85,SmithGSH85}.

A number of experiments support a Gaussian density of states $D(B,E)$ model
for two-dimensional electron systems \cite{WildeSHH06,ZhuUMP03,WangTSS92,GornikLSS85}. 
The density of states at an energy $E$ in its Gaussian form is given as
\begin{equation}
D(B,E) = \frac{e B}{ \pi \hbar} \sum_n \frac{1}{\sqrt{2 \pi}\Gamma}\; 
\exp\left[-\frac{(E - E_n)^2}{ 2 \Gamma^2 } \right]\;,
\label{eq:GaussianDOS}
\end{equation} 
where $e$ is the electron charge, $\hbar = h/2\pi$ and $h$ is Planck's constant. 
The $n$th Landau level is given as  $E_n = (n + \frac{1}{2}) \hbar \omega_c$ 
where $\omega_c = e B/m^*$ is the cyclotron frequency.  The effective mass
$m^{*}$ used here is that of carriers in a GaAs/AlGaAs heterostructure which is equal to 
$0.0667\;m_e$, where $m_e$ is the mass of the electron. 
Some experimental data fits  use  Eq.\ (\ref{eq:GaussianDOS}) with
an additional constant background obtained from a fraction of the zero-field DOS 
\cite{WildeSHH06,ZhuUMP03}. This  fraction provides an adjustable parameter dependent on experimental factors. 
However, depending on $B$,  Eq.\ (\ref{eq:GaussianDOS}) is found to 
produce Gaussian peaks with overlap between Landau levels. This overlap acts as a sort of a uniform background. Therefore, there is no need to introduce an additional term to 
Eq.\ (\ref{eq:GaussianDOS}).

In our simulations, we investigate the case when the broadening parameter 
$\Gamma$ is set to (i) fixed values, (ii) a function of 
$\gamma B^{1/2}$ where $\gamma = 0.01$ meV/T$^{1/2}$, 
and  (iii) an oscillating function of the filling factor $\nu$ similar to Ref.~\cite{WangTSS92}
such that
\begin{equation}
\Gamma(B)=0.704B^{1/2}+0.296B^{3/2}[1.8\cos^6(2\pi\nu)-1]\;,
\end{equation}
where $\nu=hN/eB$.  
For constant $\Gamma$, only the results for $\Gamma =0.2$ meV and $1.0$ meV will be shown.

\begin{figure}
    \centering 
    \includegraphics[width=\columnwidth,height=!]{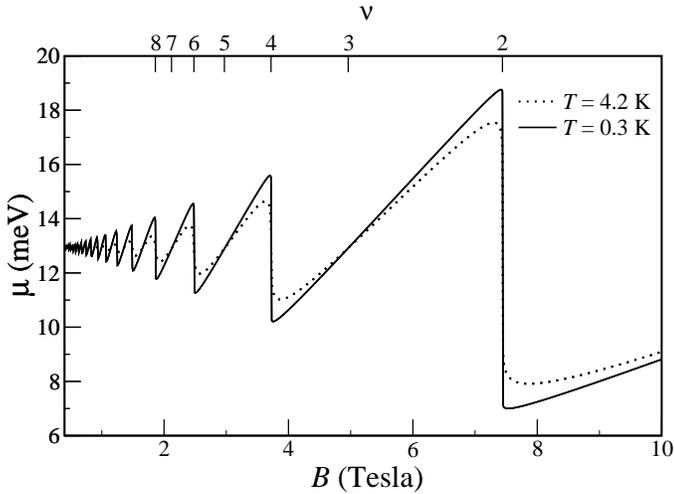}
    \caption{The chemical potential with
    broadened levels using $\Gamma = 0.2$ meV at different temperatures.
    The Landau levels are located on the diagonal 
    slopes of the saw-tooth like oscillations.}
    \label{fig:mu}
 \end{figure}

\section{Results and Discussion}

In the absence of Landau level broadening, the chemical potential exhibits 
oscillations as a function of $B$ with sharp peaks occuring at even filling 
factors $\nu$.
Since the last occupied Landau level is fullly filled for even $\nu$, 
this implies that one needs to go to the next higher level to add one more electron.
On the contrary, an odd $\nu$ corresponds to a last occupied level filled to half 
its degeneracy. Adding an electron will not change $\mu$.
Therefore, at odd $\nu$ the chemical potential is a constant.

The sharp oscillations of $\mu(B,T)$
are reduced and softened  upon the introduction of broadening
 as shown in Fig.~\ref{fig:mu} for $T=4.2$ K at $\Gamma = 0.2$ meV. 
A similar effect was observed in 2DES in quantum wells for multiple 
sub-band occupation \cite{OzdemirYO04}. 
Note, however, that despite  the presence of broadening, $\mu$
approaches its ideal behavior of having  sharp peaks  at  $T=0.3$ K.

\begin{figure}
    \centering 
    \includegraphics[width=\columnwidth,height=!]{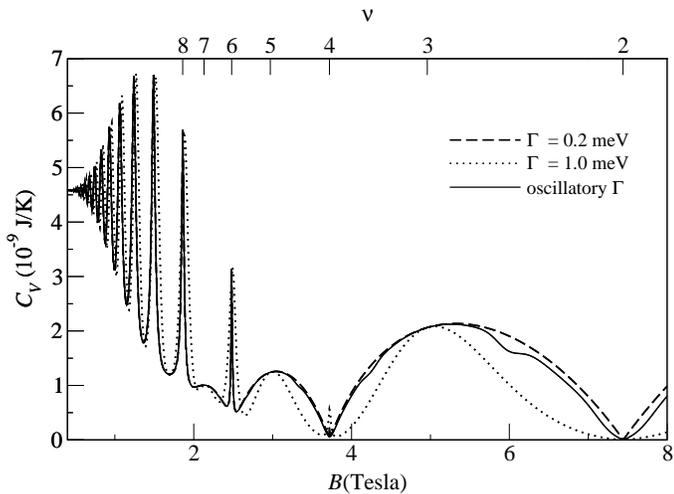}
    \caption{
The specific heat oscillations at constant $T=4.2$ K for different 
broadening parameters.}
 \label{fig:sphtgamma}
 \end{figure}

The Landau level broadening is reflected in the specific heat
in Fig.~\ref{fig:sphtgamma} as spikes at the low field region.
At $\nu > 4$, the narrow peaks occur at magnetic fields when the
chemical potential crosses from one Landau level to the next.
These crossings are known as interlevel excitations \cite{MacDonaldOL86,ZawadzkiL84}. 
As $B$ increases, the Landau levels are farther apart and it
becomes more difficult to have interlevel excitations.
This is reflected in the diminishing peak heights of $C_V$ at $\nu=8$ and 
$\nu=6$ in Fig.~\ref{fig:sphtgamma}. 
Until finally, $C_V$ vanishes at $\nu=2$.
This is comparable with the results in Ref.~\cite{XieLD90}
where interlevel excitations appear as small structures near the minima in the 
$C_V$ oscillations at high $B$.

At high $B$ ($\nu \leq 4$), Landau levels are far apart and
 the degree of degeneracy within a broadened level is high. 
Excitations brought about by an increase in $B$ are limited within the 
last occupied level. In this $B$ region the intralevel contributions 
dominate \cite{MacDonaldOL86,ZawadzkiL84}.
For example, $C_V$ exhibits a wide peak at $2 < \nu < 4$.

This general trend of competing dominance between interlevel and
intralevel contributions of $C_V$ with $B$ presented in Fig.~\ref{fig:sphtgamma}
is conserved regardless of the broadening parameter used. 
An increase in magnitude of the 
constant $\Gamma$ appears in $C_V$ as widening of the interlevel contributions 
and narrowing of the intralevel excitations. 
The latter is easily observed in the region $2 < \nu < 4$
wherein the change of $C_V$ has a slower slope with $B$. This is consistent
with the fact that a wider Landau level corresponds to a 
slower variation of the DOS with energy.

When $\Gamma$ oscillates with $B$ one obtains a similar trend for $C_V$ except for
slight variations in the region where intralevel contributions dominate.
Although not shown here, the $C_V$ result for $\Gamma=\gamma B^{1/2}$ coincides 
with the result for $\Gamma = 0.2$ meV at $T=4.2$ K.

\begin{figure}
    \centering 
    \includegraphics[width=\columnwidth,height=!]{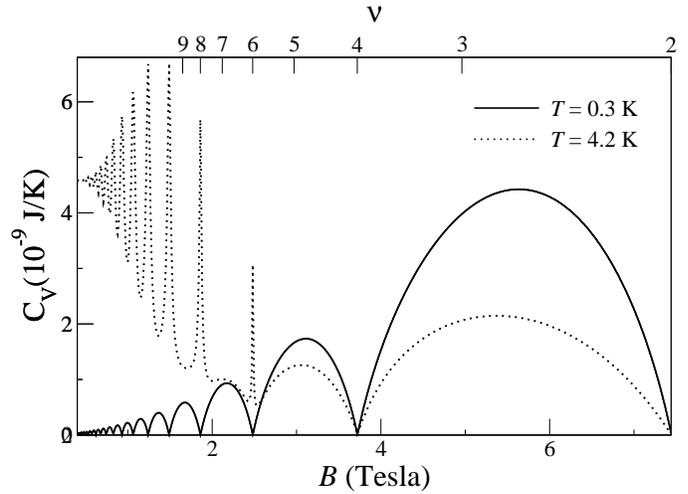}
    \caption{The heat capacity for $\Gamma=0.2$ meV at two different temperatures.}
    \label{fig:spht2T}
 \end{figure}

The influence of the temperature on the specific heat as $T\rightarrow 0$
is illustrated in Fig.~\ref{fig:spht2T}. For clarity, we will only show here
the result for $T=4.2$ K and $T=0.3$ K.
Reducing $T$ from 4.2 K minimizes the interlevel contributions
and, consequently, the magnitude of the sharp peaks at $B<4$ T diminishes. 
Until at $T=0.3$ K, the effect of $\Gamma$ vanishes. 
This shows that despite the presence of broadened Landau levels 
excitations between them are no longer possible.
Only intralevel excitations contribute to $C_V$ which is a characteristic
of an ideal 2DES.
At even $\nu$, where $\nu \geq 6$ for $\Gamma = 0.2$ meV,
 $C_V$ changes from a maximum when $T=4.2$K to a minimum when $T=0.3$ K.
This crossover is likewise found here for $\Gamma = 1.0$ meV at $\nu \geq 4$ 
and in Ref.~\cite{MacDonaldOL86} for the case when $\Gamma \propto B^{1/2}$.

\begin{figure}
    \centering 
    \includegraphics[width=\columnwidth,height=!]{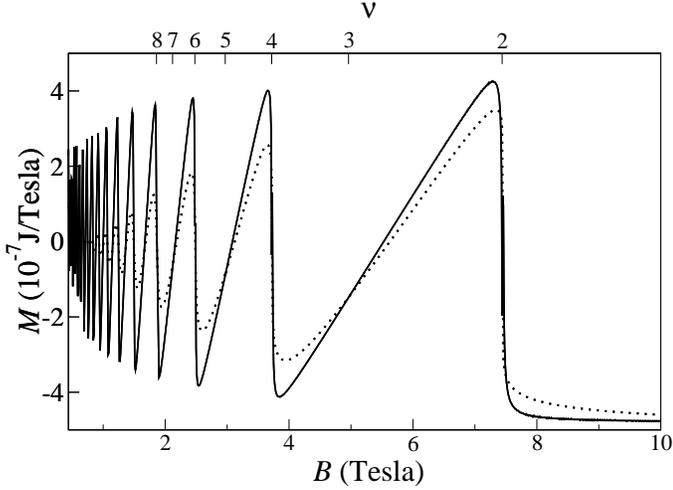}
    \caption{The magnetization for $\Gamma=0.2$ meV at $T=4.2$ K for the 
     dashed curve and at $T=0.3$ K for the solid curve.}
    \label{fig:mag}
 \end{figure}

The corresponding influence of $T$ on $M$ is shown in Fig.~\ref{fig:mag}. 
Orbital magnetization in 2DES with electron-impurity scattering 
do not have sharp spikes with $B$ according to a self-consistent theory 
\cite{XieLD90}.
Moreover, experimental evidence of the de Haas-van Alphen effect
displays rounding at the extrema of the $M$ oscillations which are attributed 
to finite temperature and to having no states in the energy gaps \cite{WildeSHH06}.
The numerical calculations presented herein, likewise, observe
these rounding of $M$ oscillations at $T=4.2$ K. 
Similar to the experimental results \cite{WildeSHH06}, the amplitude of the $M$ oscillations approaches the ideal saw-tooth behavior with respect to $B$ at  $T=0.3$ K. 
The important result here is that these sharp oscillations
are observed regardless of the form of the broadening, even as large as $\Gamma = 1.0$ meV
where there are significant amount of states in between Landau levels.

\begin{figure}
    \centering 
    \includegraphics[width=\columnwidth,height=!]{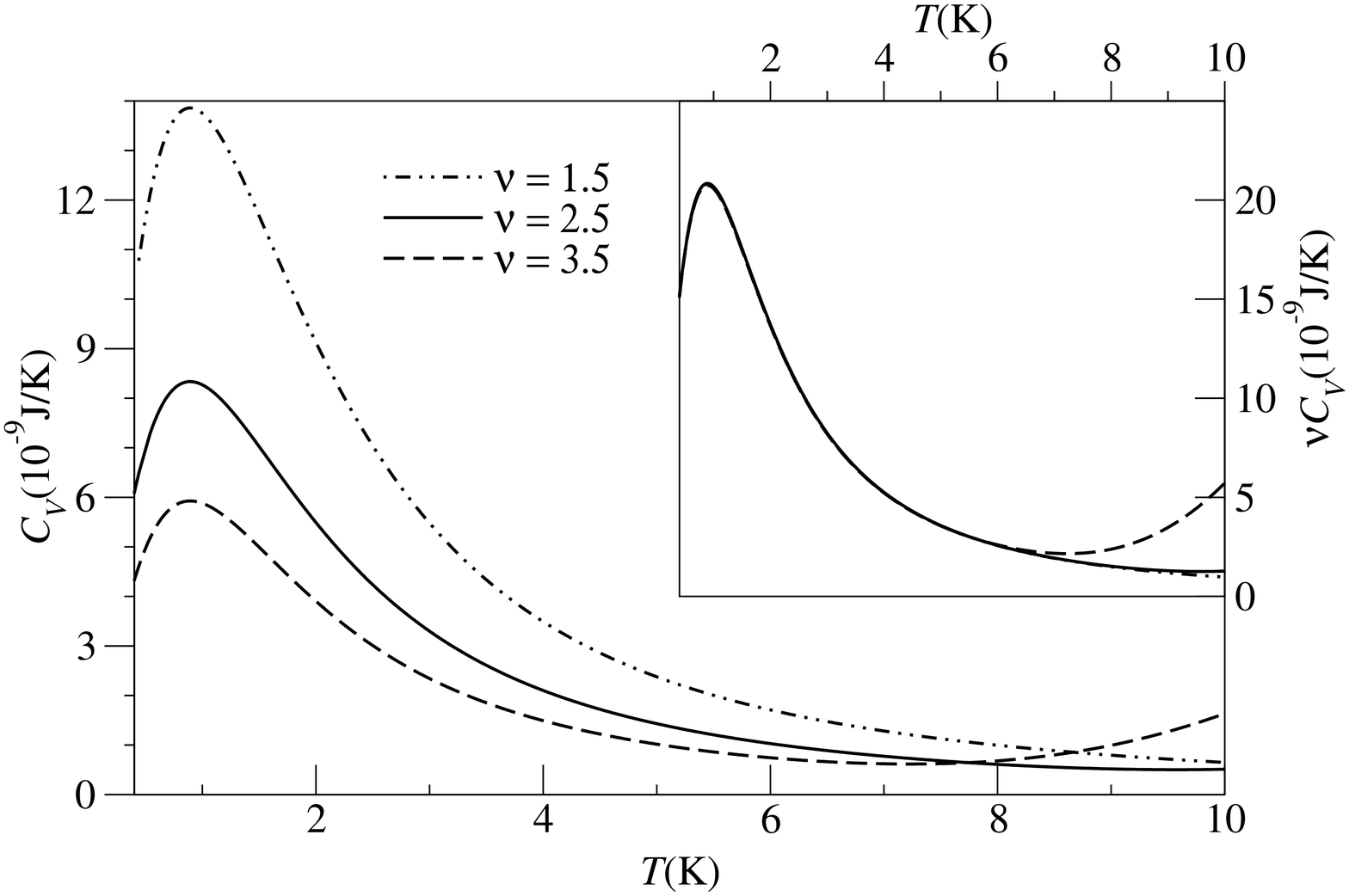}
   \caption{The temperature dependence of the specific heat  at different
	filling factors for $\Gamma = 0.2$ meV. The inset shows the
	three $C_V$ curves scaled with respect to the filling factor.}
    \label{fig:scaled}
 \end{figure}

The temperature behavior of $C_V$ corresponding to different fractional filling 
factors  is illustrated in Fig.~\ref{fig:scaled}. A peak is observed at $T_p=0.89$ K.
This characteristic temperature signifies the shift from an ideal 2DES behavior
to the regime where interlevel excitations and the Landau level broadening are felt.

Below $T_p$, $C_V$ has a linear $T$-dependence reminiscent of the bulk electron 
gas behavior \cite{AshcroftM76}. Right above $T_p$, $C_V$ decays since
the intralevel contributions decrease with $T$ for a given $B$. 
A similar peak was observed in Ref.~\cite{XieLD90}.
But they did not observe a uniform peak since their Landau level broadening has a 
non-monotonic dependence with $B$ and $N$. Consequently, their DOS may 
overlap below the Fermi energy and have large excitation gaps beyond.
The work presented here does not include spin splitting. It is worthwhile
to mention that, when spin textures are present, $C_V$ peak shifts to higher $T$ with the
Land\'{e} $g$ factor for a given fractional $\nu$ \cite{ChakrabortyP97}.

The same characteristic temperature is observed regardless of the filling factor
and the location of the chemical potential in the DOS.
Interestingly,  the $C_V$ curves for the fractional $\nu$ coincide if one scales 
the specific heat with respect to the filling factor as shown in the inset 
of Fig.~\ref{fig:scaled}. This $C_V$ behavior having a peak at $T_p$ is also observed 
at integer $\nu$. However, they do not follow this scaling behavior since at odd $\nu$, the 
chemical potential is a constant. At even $\nu$, the $C_V$ peak 
varies in magnitude depending on whether the interlevel contributions dominate or not.

\section{Conclusions}

The effect of Landau level broadening and temperature is considered in two-dimensional
electron systems using a Gaussian-shaped density of states. The broadening parameters
considered took different forms such as different constant values, a square-root dependence 
on the magnetic field and an oscillating function with respect to the filling factor.
The observed $B$ behavior of the chemical potential, specific heat and magnetization
show the same trend regardless of the form of $\Gamma$. The broadening only plays a role 
in the  width of the oscillations with $B$ for each interlevel and intralevel contribution.
A characteristic temperature is observed such that the 2DES approaches its ideal electron
gas behavior. Below this characteristic temperature, the Landau level
broadening no longer affects the behavior of the thermodynamics properties.
This indicates that scattering mechanisms present in 2DES are suppressed below $T_p$.
The magnitude of $T_p$ is obtained here using the electron concentration and 
effective mass value for a GaAs system. It would be interesting to know if experiments
are able to measure $T_p$. If not, this should be a ground for theoretical studies 
to find out why $T_p$ is masked in experiments.

\section*{Acknowledgment}
Part of the calculations were carried out at the High Performance Computing 
Facility of the Computational Science Research Center, University of the 
Philippines, which is funded by the Philippine Council 
for Advanced Science and Technology Research and Development.

\nocite{XieLD90}


\bibliographystyle{plain}

\end{document}